\documentclass{revtex4-2}
\usepackage[utf8]{inputenc}
\usepackage{amsbsy}
\usepackage{amsopn}
\usepackage{amstext}
\usepackage{amsmath,amsthm,amsfonts,amssymb}
\usepackage[mathcal]{eucal}
\usepackage{mathrsfs}
\usepackage{braket}
\usepackage[english]{babel}
\usepackage{color}
\usepackage{esint}
\usepackage{booktabs}
\usepackage{hologo}
\usepackage{graphicx}
\usepackage{float}
\usepackage{units}
\usepackage{textcomp}
\usepackage{orcidlink}

\DeclareGraphicsExtensions{.png,PNG,.pdf,.PDF}
\usepackage{hyperref}
\usepackage{slashed}
\newcommand{\ie}{\begin{equation}}
\newcommand{\fe}{\end{equation}}
\newcommand{\se}{\begin{eqnarray}}
\newcommand{\ff}{\end{eqnarray}}
\begin{document}

\title{Relativistic and nonrelativistic Landau levels for Dirac fermions in the cosmic string spacetime in the context of rainbow gravity}
\author{R. R. S. Oliveira\,\orcidlink{0000-0002-6346-0720}}
\email{rubensrso@fisica.ufc.br}
\affiliation{Departamento de F\'isica, Universidade Federal do Cear\'a (UFC), Campus do Pici, C.P. 6030, Fortaleza, CE, 60455-760, Brazil}


\date{\today}

\begin{abstract}

In this paper, we study the relativistic and nonrelativistic Landau levels for Dirac fermions in the cosmic string spacetime in the context of rainbow gravity in $(3+1)$-dimensions, where we work with the curved Dirac equation with minimal coupling in cylindrical coordinates. Using the tetrads formalism of General Relativity, we obtain a second-order differential equation. Solving this differential equation, we obtain a generalized Laguerre equation as well as the relativistic Landau levels for the fermion and antifermion, where such energy levels are quantized in terms of the quantum numbers $n$, $m_j$ and $m_s$, and explicitly depends on the rainbow functions $f(\varepsilon)$ and $g(\varepsilon)$, charge parameter $\sigma$, cyclotron frequency $\omega_c$, curvature parameter $\alpha$, and on the square rest energy $m^2_0$ and square $z$-momentum $k^2_z$, respectively. Posteriorly, we study the nonrelativistic limit of the system, where we obtain the nonrelativistic Landau levels. In both cases (relativistic and nonrelativistic), we graphically analyze the behavior of Landau levels for the three rainbow gravity scenarios as a function of the magnetic field $B$ and of the curvature parameter $\alpha$. In addition, we also compared our problem with other works, where we verified that our results generalize several particular cases in the literature.

\end{abstract}

\maketitle

\section{Introduction}

Originally arising in nonrelativistic quantum mechanics (NRQM), the so-called Landau levels (named after Lev Landau) are basically a product (result) of so-called Landau quantization, which refers to the quantization of the cyclotron orbits of charged particles in a uniform external magnetic field \cite{Landau}. In other words, the charged particles in such magnetic fields can only occupy orbits with discrete energy values, so-called Landau levels (i.e., are the quantized energy levels of charged particles in uniform magnetic fields). So, in addition to being extensively studied in NRQM \cite{Bhuiyan,Li1999,Vagner,Wakamatsu1,Wakamatsu2,Ribeiro,Rosas,Johnson}, and in relativistic quantum mechanics (RQM) \cite{Johnson,Haldane,Schakel,Lamata,Miransky,Ishikawa,Bermudez,Jentschura,Naseri}, the Landau levels was also studied in several different physical contexts, such as in condensed matter Weyl systems \cite{Laurila,Arjona}, gravitational fields (i.e., Kerr and Schwarzschild spacetimes) \cite{Landry,Hammad}, Dirac oscillator \cite{Oliveira1,Guvendi,Cunha2}, Aharonov-Bohm-Coulomb system \cite{Oliveira2}, rotating frames (inertial effects) \cite{Brand,Chen,Mameda,Liu}, neutron stars \cite{Gao}, persistent currents \cite{Kim}, surface magnetic catalysis \cite{Chen2017}, minimal length scenario \cite{Menculini}, quantum chromodynamics (QCD) \cite{Bruckmann}, etc. Furthermore, the Landau levels have already been observed in graphene \cite{Taychatanapat}, graphite \cite{Li}, semiconductors \cite{Bordacs}, Weyl semimetals \cite{Yuan}, etc; and is a central or key ingredient to explain the integer quantum Hall effect \cite{Klitzing,Gusynin}. Recently, the Landau levels were analyzed in the noncommutative quantum Hall effect with anomalous magnetic moment in different relativistic scenarios \cite{Oliveira3,GRG2024}.

In General Relativity (GR), the so-called cosmic strings are a type of linear gravitational topological defect (still hypothetical) that likely arose due to cosmological phase transitions with rotational symmetry breaking (or spontaneous gauge symmetry breaking) during the initial stages of the early universe \cite{Kibble,Vilenkin1,Vilenkin3,Bezerra,Mazur}. In particular, the cosmic strings are highly dense relativistic objects, stable, infinitely long and straight (are unidimensional defects), can be static or rotating (spinning), have cylindrical symmetry, are one of the exactly solvable models of GR, and the spacetime generated by them presents a locally flat geometry (but not globally) with a conical singularity (or non-trivial conical topology) at the origin characterized by a planar angular deficit, that is, the total angle around the cosmic string is lower than 360º (therefore, cosmic strings are also conical defects with positive curvature) \cite{Kibble,Vilenkin1,Vilenkin3,Bezerra,Mazur,Oliveira2019,OliveiraEPJC,INDIANA}. From an observational point of view, some detectors will try to look for cosmic string signals, such as the Laser Interferometer Gravitational-Wave Observatory (LIGO) \cite{Cui}, the Laser Interferometer Space Antenna (LISA) \cite{Auclair}, and the North American Nanohertz Observatory for Gravitational Waves (NANOGrav) \cite{Blasi}. Now, from the point of view of condensed matter or solid-state physics, there is a type of defect, so-called disclinations, which appears in liquid crystals and crystalline solids \cite{Kleman,L,Katanaev}, which have some interesting similarities with the cosmic strings, i.e., they are also linear topological defects with a conical singularity and break rotational symmetry of the medium \cite{Mello}. Furthermore, some notable works that analyzed the relativistic and nonrelativistic Landau levels in the cosmic string spacetime or in the presence of a disclination, we can indicate Refs. \cite{Marques,Oliveira,Muniz,Medeiros,Cunha}.

Still, in the context of GR, the so-called rainbow gravity (or gravity’s rainbow) is a semi-classical approach/attempt to probe high-energy phenomena of quantum gravity via the energy-momentum relativistic dispersion relation with higher-order terms originated (introduced) by the so-called rainbow functions, thus implying in a Lorentz symmetry violation on this energy scale \cite{Magueijo1,Magueijo2,Magueijo3,Amelino1,Amelino2,Bezerra2019,Bakke2018}. In particular, these rainbow functions explicitly depend on the ratio between the energy of a test particle (i.g., bosons or fermions) and on the Planck energy (i.e., $\varepsilon=E/E_P\leq 1$); consequently, this will govern departures from usual relativistic Minkowskian expressions. In this way, an interesting phenomenological framework to study the quantum gravity effects is via Doubly General Relativity (Nonlinear or Doubly Special Relativity in GR) or, as it is mostly known, via rainbow gravity \cite{Magueijo1,Magueijo2,Magueijo3,Amelino1,Amelino2,Bezerra2019,Bakke2018}. Here, the terminology ``doubly special'', or ``doubly general'', is a very suggestive term since in rainbow gravity there are two quantities that are observer-independent (invariant quantities), which are the speed of light and the Planck energy (or Planck length, since for $\hbar=c=1$, so $E_P=L_P^{-1}=const.>0$) \cite{Magueijo1,Magueijo2,Magueijo3,Amelino1,Amelino2,Bezerra2019,Bakke2018}. That is, now $E_P$ is also a universal constant for all inertial observers. Therefore, in order to build this framework, the line element (or metric) of spacetime has also to be energy-dependent (a consequence of a nonlinear Lorentz transformation in momentum space) in such a way that this dependence becomes stronger as the energy of the probe particle approaches the Planck energy/scale (i.e., $\varepsilon\to 1$) \cite{Magueijo1,Magueijo2,Magueijo3,Amelino1,Amelino2,Bezerra2019,Bakke2018}.

So, unlike the usual GR, where gravity acts equally regardless of the energy of the particles, rainbow gravity acts differently on particles with different energies, that is, for each energy value of the particle, it feels different gravity levels (therefore, the terminology ``rainbow'' is also a very suggestive term since rainbow gravity affects different energies, wavelengths, or frequencies associated with the particle in the same way that a prism affects light). Indeed, this effect is very small for objects like the Earth (where Newtonian gravity acts), however, it should become significant for objects like black holes (where GR or ``quantum gravity'' acts). On the other hand, another ``simple" way to detect such an effect would be, for example, with ultra-high energy cosmic rays or TeV photons and neutrinos from gamma-ray-bursts (GRB), where the relativistic dispersion relation should acquire corrections (modifications) due to the parameter $\varepsilon$ \cite{Magueijo2,AmelinoCamelia1,AmelinoCamelia2,Jacob,Zhang,Nilsson}. Thus, after being introduced into the literature, rainbow gravity gained a lot of attention and also many interesting applications, such as in neutron stars \cite{Hendi1,Panah}, black holes \cite{Hendi2,Hendi3,Gim,Ling,Ali1,Ali2}, wormholes \cite{Amirabi}, Friedmann-Robertson-Walker (FRW) cosmology \cite{Hendi,Awad}, Casimir effect \cite{Bezerra2017}, Bose-Einstein condensates \cite{Furtado}, Klein–Gordon oscillator \cite{Montigny1}, global monopoles \cite{Faizuddin}, etc. Later, a more ``general'' theory was introduced into the literature for combining cosmic strings with rainbow gravity (i.e., a ``cosmic rainbow gravity'') \cite{Momeni}. For example, considering problems in NRQM and RQM, such a combination has already been applied in the Landau levels (only for spinless particles) \cite{Bezerra2019}, Dirac, Klein-Gordon and  Duffin–Kemmer–Petiau oscillators \cite{Bakke2018,Santos,Hosseinpour}, Aharonov–Bohm effect \cite{B1}, Landau–Aharonov–Casher quantization \cite{B2}, wormholes \cite{Ahmed}, electrostatic self-interaction \cite{Santos2019}, bosons with position-dependent mass \cite{Mustafa}, etc.

This paper has as its goal to study the relativistic and nonrelativistic Landau levels for Dirac fermions (spin-1/2 particles) in the cosmic string spacetime in the context of rainbow gravity in $(3+1)$-dimensions (i.e., we will study the case or fermionic version of Ref. \cite{Bezerra2019}). In other words, here we will do a study within the context of rainbow gravity by means of the effects that stem from high-order corrections to the relativistic dispersion relation for the fermionic Landau levels (energy spectrum) in the static cosmic string spacetime. So, to carry out such a study, we will work with the curved Dirac equation with minimal coupling in cylindrical coordinates, where the formalism used will be the tetrads or vierbein formalism of GR. In fact, we chose this formalism because it is considered an excellent tool for studying fermions in curved spacetimes. Furthermore, we also consider a gauge field provided by a uniform external magnetic field in the same direction (or symmetry axis) of the cosmic string (i.e., $z$-axis). In particular, the cosmic string spacetime is defined (introduced) through a change in the angular coordinate of the system as $\varphi\to\alpha\varphi$, where $\alpha\equiv 1-4G\mu/c^2$ ($0<\alpha\leq 1$) is a curvature or topological parameter, which generates an angle deficit given by $\delta=2\pi(1-\alpha)=8\pi G\mu/c^2$, with $\mu\geq 0$ being the linear mass density (mass per unit length) of the cosmic string, $G$ is Newton's constant of gravitation, and $c$ is the speed of light, respectively \cite{Bezerra,Mazur,Oliveira2019,INDIANA,OliveiraEPJC,Oliveira,Muniz,Medeiros,Cunha}. From a geometric point of view, the cosmic string spacetime corresponds to a Minkowski spacetime with a conical singularity at the origin (i.e., ``a cosmic string cuts the flat spacetime in the shape of a wedge by an angle $\delta$''). In fact, in the absence of the cosmic string where $\alpha=1$ (or $\mu=0$), we recover the Minkowski spacetime (as it should be). Now, with respect to rainbow functions, in this paper, we will consider three different pairs of rainbow functions for $f(\varepsilon)$ and $g(\varepsilon)$, whose choice is based on phenomenological motivations (and most worked in the literature). In this way, we will work in three different scenarios (or three different cases), given as follows:
\begin{itemize}
\item First Scenario. The following functions were studied in Refs. \cite{Ali2,Amelino1,Amelino2,Bezerra2019,Hendi2,Hendi3,Gim,Furtado,Hendi,Hendi1,Panah,Mustafa}:
\begin{equation}\label{f1}
f(\varepsilon)=1, \ \ g(\varepsilon)=\sqrt{1-\eta\varepsilon},
\end{equation}
which is compatible with results from Loop Quantum Gravity and noncommutative spacetime, and $\eta>0$ is a dimensionless constant (free) parameter (of order one), and $\varepsilon$ is sometimes called the rainbow parameter (a very suggestive name by the way).
\item Second Scenario. The following functions were studied in Refs. \cite{Ali2,AmelinoCamelia1,Bezerra2019,Hendi2,Hendi3,Ling,Amirabi,Bezerra2017,Furtado,Hosseinpour,Santos2019,Hendi,Awad,Hendi1,Panah,Mustafa}:
\begin{equation}\label{f2}
f(\varepsilon)=\frac{e^{\eta\varepsilon}-1}{\eta\varepsilon}, \ \ g(\varepsilon)=1,
\end{equation}
which is compatible with the hard spectra from gamma-ray bursters at cosmological distances.
\item Third Scenario. The following functions were studied in Refs. \cite{Ali2,Magueijo1,Bakke2018,Bezerra2019,Magueijo3,Hendi2,Hendi3,Amirabi,Bezerra2017,Furtado,Santos,Hosseinpour,Hendi,Awad,Hendi1,Panah,Mustafa}:
\begin{equation}\label{f3}
f(\varepsilon)=g(\varepsilon)=\frac{1}{1-\eta\varepsilon},
\end{equation}
which produces a constant speed of light, and might
solve the horizon problem.
\end{itemize}

The structure of this paper is organized as follows. In Sect. \ref{sec2}, we introduce the curved Dirac equation with minimal coupling in the cosmic string spacetime in the context of rainbow gravity. Using the tetrads formalism (or spin connection), we obtain a second-order differential equation. In Sect. \ref{sec3}, we solve exactly and analytically this differential equation through a new independent variable and of the asymptotic behavior. Done this, we obtain a generalized Laguerre equation, where the last term of this equation obeys a quantization condition (i.e., the last term must be equal to a negative integer). Consequently, we obtain from this condition the relativistic Landau levels (relativistic energy spectrum) for a fermion and an antifermion. In this section, we graphically analyze the behavior of Landau levels for the three scenarios as a function of the magnetic field and of the parameter $\alpha$. In Sect. \ref{sec4}, we study the nonrelativistic limit (low-energy limit) of our results, where we obtain the nonrelativistic Landau levels (nonrelativistic energy spectrum). In this section, we also graphically analyze the behavior of Landau levels for the three scenarios as a function of the magnetic field and of the parameter $\alpha$. Besides, in both cases (relativistic and nonrelativistic), we also compared our problem with other works, where we verified that our results generalize several particular cases in the literature. Finally, in Sect. \ref{sec5} we present our conclusions. Here, we use the natural units $(\hslash=c=G=1)$ and the spacetime with signature $(+,-,-,-)$.

\section{The Dirac equation in the cosmic string spacetime in the context of rainbow gravity \label{sec2}}

The $(3+1)$-dimensional covariant Dirac equation with minimal coupling in a generic curved spacetime is given by the following expression (in cylindrical coordinates) \cite{Oliveira3,GRG2024,Greiner,Lawrie,Chen,Cunha2}
\begin{equation}\label{dirac1}
\left\{i\gamma^\mu(x)[D_\mu(x)+\Gamma_\mu (x)]-m_0\right\}\psi(t,r,\theta,z)=0, \ \ (\mu=t,r,\theta,z),
\end{equation}
where $\gamma^{\mu}(x)=e^\mu_{\ a}(x)\gamma^a$ are the curved gamma matrices, which satisfy
the anticommutation relation of the covariant Clifford algebra: $\{\gamma^{\mu}(x),\gamma^{\nu}(x)\}=2g^{\mu\nu}(x)$, being $g^{\mu\nu}(x)$ the curved metric tensor (or curved metric), $\gamma^a=(\gamma^0,\gamma^i)=(\gamma^0,\gamma^1,\gamma^2,\gamma^3)=(\gamma_0,-\gamma_1,-\gamma_2,-\gamma_3)=\eta^{ab}\gamma_b$ are the usual or flat gamma matrices (in Cartesian coordinates), $\eta^{ab}=\eta_{ab}=$diag$(+1,-1,-1,-1)$ is the flat or Minkokski metric, $e^\mu_{\ a}(x)$ are the tetrads (tetrad fields) or vierbein, $D_\mu(x)=\partial_\mu+iqA_\mu (x)$ is the covariant derivative (or minimal coupling), being $\partial_\mu=(\partial_t,\partial_r,\partial_\theta,\partial_z)$ the usual partial derivatives, $\Gamma_\mu(x)=-\frac{i}{4}\omega_{ab\mu}(x)\sigma^{ab}$ is the spinorial connection (or spinor affine connection), being $\omega_{ab\mu}(x)$ the spin connection, $\sigma^{ab}=\frac{i}{2}[\gamma^a,\gamma^b]=i\gamma^a\gamma^b$ ($a\neq b$) is a flat antisymmetric tensor, $A_\mu (x)=e^b_{\ \mu}(x)A_{b}$ is the curved electromagnetic four-potential (or curved gauge field), $e^b_{\ \mu}(x)$ are the inverse tetrads, $A_b=(0,0,A_2,0)=(0,0,-A_\theta,0)$ is the flat electromagnetic four-potential (or flat gauge field), $\psi=\ket{\psi}=e^{\frac{i\theta\Sigma^3}{2}}\Psi_D$ is the four-component curvilinear spinor, where $\Psi_D\in\mathbb{C}^4$ is the original Dirac spinor (four-element column vector or ``relativistic spin-1/2 wave function''), and the physical quantities $q=\pm e$, and $m_0>0$, are the electric charge (coupling constant), and the rest mass of the Dirac fermion, respectively. Here, the Latin indices $(a, b, c, \ldots)$ are used to label the local coordinates system (local reference frame or the Minkowski spacetime) while the Greek indices $(\mu, \nu, \alpha, \ldots)$ are used for the general coordinates system (general reference frame or the curved spacetime).

Explicitly, the flat gamma matrices are given as follows (in the standard or Dirac representation)
\begin{equation}\label{Diracmatrices}
 \gamma^0=\left(
    \begin{array}{cc}
      I & \ 0 \\
      0 & -I \\
    \end{array}
  \right), \ \ \gamma^i=\left(
    \begin{array}{cc}
      0\ &  \sigma^i \\
      -\sigma^i\ & 0 \\
    \end{array}
  \right), \ \ (i=1,2,3=x,y,z),
\end{equation}
where $I$ is a $2\times 2$ unit matrix and $\sigma^i=\eta^{ij}\sigma_j$ are the
standard Pauli matrices ($2\times 2$ matrices), written as 
\begin{equation}\label{Paulimatrices}
\sigma_1=\left(
    \begin{array}{cc}
      0\ &  1 \\
      1\ & 0 \\
    \end{array}
  \right), \ \  \sigma_2=\left(
    \begin{array}{cc}
      0 & -i  \\
      i & \ 0 \\
    \end{array}
  \right), \ \  \sigma_3=\left(
    \begin{array}{cc}
      1 & \ 0 \\
      0 & -1 \\
    \end{array}
  \right),
\end{equation}
and satisfies $\{\sigma^i,\sigma^j\}=2\delta^{ij}$, $[\sigma^i,\sigma^j]=2i\epsilon^{ijk}\sigma^k$, and $(\sigma^i)^2=I=\delta^{ii}$.

In addition, the spin connection (an antisymmetric tensor in its internal indices $a$ and $b$) is defined in the following form (torsion-free) \cite{Lawrie}
\begin{equation}\label{spinconnection}
\omega_{ab\mu}(x)=-\omega_{ba\mu}(x)=\eta_{ac}e^c_{\ \nu}(x)\left[e^\sigma_{\ b}(x)\Gamma^\nu_{\ \mu\sigma}(x)+\partial_\mu e^\nu_{\ b}(x)\right], 
\end{equation}
where $\Gamma^\nu_{\ \mu\sigma}(x)$ are the Christoffel symbols of the second type (a symmetric tensor), and written as
\begin{equation}\label{Christoffel}
\Gamma^\nu_{\ \mu\sigma}(x)=\frac{1}{2}g^{\nu\lambda}(x)\left[\partial_\mu g_{\lambda\sigma}(x)+\partial_\sigma g_{\lambda\mu}(x)-\partial_\lambda g_{\mu\sigma}(x)\right]. 
\end{equation}

On the other hand, the tetrads and their inverses must satisfy the following relations
\begin{eqnarray}\label{tetrads}
&& g_{\mu\nu}(x)=e^a_{\ \mu}(x)e^b_{\ \nu}(x)\eta_{ab},
\nonumber\\
&& g^{\mu\nu}(x)=e^\mu_{\ a}(x)e^\nu_{\ b}(x)\eta^{ab},
\nonumber\\
&& g^{\mu\sigma}(x)g_{\nu\sigma}(x)=\delta^\mu_{\ \nu}=e^a_{\ \nu}(x)e^\mu_{\ a}(x),
\end{eqnarray}
or yet
\begin{eqnarray}\label{metric3}
&& \eta_{ab}=e^\mu_{\ a}(x)e^\nu_{\ b}(x)g_{\mu\nu}(x),
\nonumber\\
&& \eta^{ab}=e^a_{\ \mu}(x)e^b_{\ \nu}(x)g^{\mu\nu}(x),
\nonumber\\
&& \eta^{ac}\eta_{cb}=\delta^a_{\ b}=e^a_{\ \mu}(x)e^\mu_{\ b}(x).
\end{eqnarray}

So, using the information above, we can rewrite Eq. \eqref{dirac1} as
\begin{eqnarray}\label{dirac2}
\left[i\gamma^t(x)\partial_t+i\gamma^r(x)\partial_r+i\gamma^\theta(x)\partial_\theta+i\gamma^z(x)\partial_z-\sigma e\gamma^{\theta}(x)A_{\theta}(x)-m_0
\right]\psi
 \nonumber \\
    +i[\gamma^t(x)\Gamma_t(x)+\gamma^{r}(x)\Gamma_{r}(x)+\gamma^{\theta}(x)\Gamma_{\theta}(x)+\gamma^{z}(x)\Gamma_{z}(x)]\psi=0,
\end{eqnarray}
where we consider $q=\pm e=\sigma e$ (i.e., we can have a positively or negatively charged fermion), and the product $\gamma^{\theta}(x)A_{\theta}(x)$ still can be written as $\gamma^\theta (x)A_\theta (x)=e^\theta_{\ 2}(x)e^2_{\ \theta}(x)\gamma^2 A_2=\delta^2_{\ 2}\gamma^2 A_2=\gamma^2 A_2=-\gamma^2A_{\theta}$ (do not confuse both potentials $A_\theta(x)$ and $A_\theta$, i.e., one will depend on $\alpha$ and $\varepsilon$ and the other does not).

Now, let us focus our attention on the line element of the cosmic string spacetime in the context of rainbow gravity as well as on the form of the metric (and its inverse), tetrads (and its inverses), curved gamma matrices, and spinorial and spin connections. Then, starting with the cosmic string spacetime, we have the following line element for such a conical curved background \cite{Bezerra,Mazur,Oliveira2019,OliveiraEPJC,INDIANA,Oliveira,Muniz,Medeiros,Cunha}
\begin{equation}\label{lineelement1}
ds^2=g_{\mu\nu}(x)dx^\mu dx^\nu=dt^2-dr^2-\alpha^2 r^2d\theta^2-dz^2, \ \ (\mu,\nu=t,r,\theta,z),
\end{equation}
where $r=\sqrt{x^2+y^2}$ ($0\leq r<\infty$) is the polar radial coordinate (radial distance or simply the radial coordinate), $0\leq\theta\leq 2\pi$ ($\theta=\tan^{-1}(x/y)$) is the angular or azimuthal coordinate, $-\infty<(t,z)<\infty$ are the temporal and axial coordinates, and $0<\alpha\leq 1$ is the curvature or topological parameter, respectively.

However, in the framework of rainbow gravity, the standard relativistic dispersion relation is modified due to the regime of the high energy scale (or Planck scale). Thereby, for a relativistic particle (probe particle) with a rest mass $m_0$ and a (canonical linear) momentum $p$, this modified relativistic dispersion relation is given as follows \cite{Magueijo1,Magueijo2,Magueijo3,Amelino1,Amelino2,Bezerra2019,Bakke2018}
\begin{equation}
E^2 f^2(\varepsilon)-p^2 g^2(\varepsilon)=m_0^2,
\end{equation}
where $\varepsilon=E/E_P=\vert E\vert/E_P$ ($0\leq\varepsilon\leq 1$), being $E$ the relativistic total energy of the probe particle and $E_P$ is the Planck energy. In particular, at high-energy scales, the rainbow functions $f(\varepsilon)$ and $g(\varepsilon)$ end up violating the standard dispersion relation since now the rainbow functions originate terms with order greater than $E^2$ in such relation. So, to describe a modified dispersion relation as a norm of the four-momentum in spacetime, the rainbow functions are absorbed in the tetrads, leading to an effective energy-dependent metric given by $g_{\mu\nu}(\varepsilon)$, or better $g_{\mu\nu}(x,\varepsilon)$ \cite{Bezerra2019}. Besides, the rainbow functions must satisfy the following conditions (limits) at a ``low-energy regime'' (infrared regime)
\begin{equation}\label{limit}
\lim_{\varepsilon\to 0} f(\varepsilon)=1, \ \ \lim_{\varepsilon\to 0} g(\varepsilon)=1,
\end{equation}
and therefore, we recover standard GR (or standard dispersion relation) in this limit (as it should be).

In that way, the framework of rainbow gravity permits us to rewrite the line element \eqref{lineelement1} in terms of the rainbow functions. Therefore, now we have the following line element of a cosmic string modified by rainbow gravity \cite{Magueijo1,Magueijo2,Magueijo3,Amelino1,Amelino2,Bezerra2019,Bakke2018}
\begin{equation}\label{lineelement2}
ds^2(\varepsilon)=g_{\mu\nu}(x,\varepsilon)dx^\mu dx^\nu=\frac{dt^2}{f^2(\varepsilon)}-\frac{1}{g^2(\varepsilon)}[dr^2+\alpha^2 r^2d\theta^2+dz^2],
\end{equation}
where $g_{\mu\nu}(x,\varepsilon)$ and $g^{\mu\nu}(x,\varepsilon)$ are defined as (obs.: the energy is not of the spacetime but of the probe particle itself)
\begin{equation}\label{metric1}
g_{\mu\nu}(x,\varepsilon)=\left(\begin{array}{cccc}
\frac{1}{f^2(\varepsilon)} & \ 0 & 0 & 0 \\
0 & -\frac{1}{g^2(\varepsilon)} &  0 & 0 \\
0 & \ 0 & -\frac{\alpha^2 r^2}{g^2(\varepsilon)} & 0 \\
0 & \ 0 & 0 & -\frac{1}{g^2(\varepsilon)}
\end{array}\right), \ \
g^{\mu\nu}(x,\varepsilon)=\left(\begin{array}{cccc}
f^2(\varepsilon) & \ 0 & 0 & 0 \\
0 & -\frac{1}{g^2(\varepsilon)} &  0 & 0 \\
0 & \ 0 & -\frac{g^2(\varepsilon)}{\alpha^2 r^2} & 0 \\
0 & \ 0 & 0 & -g^2(\varepsilon)
\end{array}\right).
\end{equation}

Note that, by taking the limit $\varepsilon\to0$, we recover the line element of the usual cosmic string spacetime ($\lim_{\varepsilon\to 0} ds^2(\varepsilon)=ds^2$). Besides, as now the spacetime metric depends on the energy of the probe particle, it implies that particles with different energies (frequencies or wavelengths) will travel on different geodesics through the spacetime, i.e., will experience different gravity levels (and the energy scale where this effect is expected to be observed is close to the Planck scale) \cite{B1}. For example, by combining Hubble + Type Ia Supernovae + Baryon Acoustic Oscillations + Cosmic Microwave Background data, the authors of Ref. \cite{Nilsson} obtained a lower bound on the energy of a massless particle which can be as high as $10^{17}$ GeV. Consequently, at this energy scale, deviations of the Lorentz invariance/symmetry should be observed (i.e., a possible rainbow gravity signature).

Thus, with the line element well defined (or metric well defined), given by the expression \eqref{lineelement2}, we now need to build a local reference frame where the observer will be placed (i.e., the laboratory frame). Consequently, it is in this local frame that we can define the gamma matrices (or the spinor) in a curved spacetime \cite{Oliveira3,GRG2024,Greiner,Chen,Cunha2}. In particular, using the tetrads formalism of GR, it is possible to achieve this objective \cite{Oliveira3,GRG2024,Greiner,Chen,Cunha2}. According to this formalism, a curved spacetime can be introduced point to point with a flat spacetime through objects of the type $e^\mu_{\ a}(x,\varepsilon)$, which are so-called tetrads (square matrices), and which together with their inverses, given by $e^a_{\ \mu}(x,\varepsilon)$, satisfy the following relations: $dx^\mu=e_{\ a}^\mu(x,\varepsilon)\hat{\theta}^a(\varepsilon)$ and $\hat{\theta}^b(\varepsilon)=e^b_{\ \nu}(x,\varepsilon)dx^\nu$, where $\hat{\theta}^a(\varepsilon)$ ($a,b=0,1,2,3$) is a quantity so-called noncoordinate basis (not to be confused with the angular coordinate $\theta$).

So, through the tetrads formalism, we can rewrite the line element \eqref{lineelement2} in terms of the noncoordinate basis, such as
\begin{equation}\label{lineelement3}
ds^2(\varepsilon)=\eta_{ab}\hat{\theta}^a(\varepsilon)\otimes\hat{\theta}^b(\varepsilon)=(\hat{\theta}^0(\varepsilon))^2-(\hat{\theta}^1(\varepsilon))^2-(\hat{\theta}^2(\varepsilon))^2-(\hat{\theta}^3(\varepsilon))^2,
\end{equation}
where the components of $\hat{\theta}^a(\varepsilon)$ are given by (i.e., a well-defined noncoordinated basis)
\begin{equation}\label{bases}
\hat{\theta}^0(\varepsilon)=\frac{1}{f(\varepsilon)}dt, \ \ \hat{\theta}^1(\varepsilon)=\frac{1}{g(\varepsilon)}dr, \ \ \hat{\theta}^2(\varepsilon)=\frac{\alpha r}{g(\varepsilon)}d\theta, \ \ \hat{\theta}^3(\varepsilon)=\frac{1}{g(\varepsilon)}dz,\ \ (dx^t=dt, \ dx^r=dr, \  dx^\theta=d\theta, \ dx^z=dz).
\end{equation}

Therefore, this implies that the tetrads and their inverses take the following form
\begin{equation}\label{tetrads}
e^{\mu}_{\ a}(x,\varepsilon)=\left(
\begin{array}{cccc}
 f(\varepsilon) & 0 & 0 & 0\\
 0 & g(\varepsilon) & 0 & 0\\
 0 & 0 & \frac{g(\varepsilon)}{\alpha r} & 0\\
 0 & 0 & 0 & g(\varepsilon)\\
\end{array}
\right), \ \
e^{a}_{\ \mu}(x,\varepsilon)=\left(
\begin{array}{cccc}
 \frac{1}{f(\varepsilon)} & 0 & 0 & 0\\
 0 & \frac{1}{g(\varepsilon)} & 0 & 0\\
 0 & 0 & \frac{\alpha r}{g(\varepsilon)} & 0\\
 0 & 0 & 0 & \frac{1}{g(\varepsilon)}\\
\end{array}
\right).
\end{equation}

Consequently, the curved gamma matrices are written as
\begin{eqnarray}\label{gammamatrices}
&& \gamma^t(x,\varepsilon)=f(\varepsilon)\gamma^0,
\nonumber\\
&& \gamma^r(x,\varepsilon)=g(\varepsilon)\gamma^1,
\nonumber\\
&& \gamma^\theta(x,\varepsilon)=\frac{g(\varepsilon)}{\alpha r}\gamma^2\\
&& \gamma^z(x,\varepsilon)=g(\varepsilon)\gamma^z.
\end{eqnarray}

Explicitly, the non-null components of the Christoffel symbols are given by
\begin{eqnarray}\label{symbols}
&& \Gamma^{r}_{\ \theta\theta}(x,\varepsilon)=-\alpha^2 r,
\nonumber\\
&& \Gamma^{\theta}_{\ r\theta}(x,\varepsilon)=\Gamma^{\theta}_{\ \theta r}(x,\varepsilon)=\frac{1}{r}.
\end{eqnarray}

Consequently, the non-null components of the spin connection are written as
\begin{equation}\label{spinconnection2}
\omega_{12\theta}(x,\varepsilon)=-\omega_{21\theta}(x,\varepsilon)=-\alpha,
\end{equation}
where implies in the following non-null component for the spinorial connection
\begin{equation}\label{spinorialconnection}
\Gamma_\theta(x,\varepsilon)=\frac{\alpha}{2}\gamma^1\gamma^2.
\end{equation}

Then, using the expressions \eqref{gammamatrices} and \eqref{spinorialconnection}, we obtain the following contribution of the spinorial connection (or spin)
\begin{equation}\label{contributionofthespinorialconnection}
\gamma^{\theta}(x,\varepsilon)\Gamma_{\theta}(x,\varepsilon)=\frac{g(\varepsilon)}{2 r}\gamma^{1}.
\end{equation}

In this way, we have from Eq. \eqref{dirac2} the following Dirac equation in the cosmic string spacetime in the context of rainbow gravity
\begin{equation}\label{dirac3}
\left[if(\varepsilon)\gamma^0\partial_t+ig(\varepsilon)\gamma^1\left(\partial_r+\frac{1}{2r}\right)+\gamma^2\left(\frac{ig(\varepsilon)}{\alpha r}\partial_\theta+\frac{m_0\sigma\omega_c}{2}r\right)+ig(\varepsilon)\gamma^3 \partial_z-m_0\right]\psi=0,
\end{equation}
where we use the fact that $\gamma^{\theta}(x,\varepsilon)A_{\theta}(x,\varepsilon)=-\frac{1}{2}Br\gamma^2$, with $A_{\theta}(x,\varepsilon)$ given by $A_{\theta}(x,\varepsilon)=\frac{\alpha r}{g(\varepsilon)}A_2=-\frac{\alpha r}{g(\varepsilon)} A_\theta=-\frac{B\alpha}{2g(\varepsilon)}r^2$ (a potential vector quadratic in the radial coordinate and dependent on $\alpha$ and $\varepsilon$) \cite{Cunha,Cunha2,Oliveira3,GRG2024,Medeiros}, being $A_\theta=\frac{1}{2}Br$ (Landau symmetric gauge) the angular component of $A_b=\eta_{bc}A^c=(A_0,A_1,A_2,A_3)=(0,0,-A_\theta,0)$ and generated through a uniform magnetic field $\Vec{B}=B\Vec{e}_z$ ($B=B_z=const.\geq 0$), and $\omega_c=\frac{eB}{m_0}\geq 0$ is the famous cyclotron frequency (an angular velocity which describes the fermion helicoidal trajector \cite{Bermudez}). In particular, the product $\gamma^\theta (x,\varepsilon)A_\theta (x,\varepsilon)$ could also be written as: $\gamma^\theta (x,\varepsilon)A_\theta (x,\varepsilon)=e^\theta_{\ 2}(x,\varepsilon)e^2_{\ \theta}(x,\varepsilon)\gamma^2 A_2=\delta^2_{\ 2}\gamma^2 A_2=\gamma^2 A_2=-\gamma^2A_{\theta}$, with $A_{\theta}=\frac{1}{2}Br$ (a potential vector linear in the radial coordinate and independent on $\alpha$ and $\varepsilon$), that is, we have the product in the Minkowski spacetime (in polar coordinates) already done a similarity transformation (this explains the existence of exponential $e^{\frac{i\theta\Sigma^3}{2}}$ in the spinor $\psi$, where such exponential has the purpose of converter $\gamma^\theta$ in $\gamma^2$ as well as $\gamma^r$ in $\gamma^1$).

In addition, here we consider a stationary quantum system; consequently, the spinor $\psi$ can be written as follows \cite{Oliveira3,GRG2024,Bakke2018,B1}
\begin{equation}\label{spinor}
\psi(t,r,\theta,z)=e^{i(m_j\theta+k_z z-Et)}\phi(r), \ \ \phi(r)=\left(
           \begin{array}{c}
            F(r) \\
            G(r) \\
           \end{array}
         \right), \ \ (F(r)\neq G(r)),
\end{equation}
where $F(r)=(F_+,F_-)^T$ and $G(r)=(G_+,G_-)^T$ are two spinors (two-component spinors each), $E$ is the relativistic total energy, $k_z=p_z=const.\geq 0$ ($-\infty<k_z<\infty$) is the module of the wavevector or momentum vector along the $z$-direction (namely, $z$-wavevector or $z$-momentum), and $m_j=\pm\frac{1}{2},\pm\frac{3}{2},\ldots$ is the total magnetic quantum number.

So, using the spinor \eqref{spinor} in \eqref{dirac3}, we have the following time-independent Dirac equation (or stationary Dirac equation)
\begin{equation}\label{dirac4}
\left[f(\varepsilon)\gamma^0 E+ig(\varepsilon)\gamma^1\left(\frac{d}{dr}+\frac{1}{2r}\right)-\gamma^2\left(\frac{g(\varepsilon)m_j}{\alpha r}-\frac{m_0\sigma\omega_c}{2}r\right)-g(\varepsilon)\gamma^3 k_z-m_0\right]\phi(r)=0.
\end{equation}

As we can see in the equation above, both quantities $g(\varepsilon)$ and $\alpha$ have the function of modifying (``shifting'') the total angular momentum of the fermion (along the $z$-direction) and, therefore, we can now say that the total magnetic quantum number will be given by $M_j(\varepsilon)=\frac{g(\varepsilon)m_j}{\alpha}$, where $J_z$ is rewritten as $J_z(\varepsilon)=-i\frac{g(\varepsilon)}{\alpha}\partial_\theta$ \cite{Bezerra}. In fact, for $g(\varepsilon)=\alpha=1$ (spacetime without cosmic string and rainbow gravity), we recover the particular or usual case \cite{Oliveira3}.

Besides, using the gamma matrices given in \eqref{Diracmatrices}, we can obtain from \eqref{dirac4} two coupled first-order differential equations for $F(r)$ and $G(r)$, where the first is given by
\begin{equation}\label{dirac5}
(m_0-f(\varepsilon)E)F(r)=\left[ig(\varepsilon)\sigma^1\left(\frac{d}{dr}+\frac{1}{2r}\right)-\sigma^2\left(\frac{g(\varepsilon)m_j}{\alpha r}-\frac{m_0\sigma\omega_c}{2}r\right)-g(\varepsilon)\sigma^3 k_z\right]G(r),
\end{equation}
while the second coupled equation is given by
\begin{equation}\label{dirac6}
(m_0+f(\varepsilon)E)G(r)=\left[-ig(\varepsilon)\sigma^1\left(\frac{d}{dr}+\frac{1}{2r}\right)+\sigma^2\left(\frac{g(\varepsilon)m_j}{\alpha r}-\frac{m_0\sigma\omega_c}{2}r\right)+g(\varepsilon)\sigma^3 k_z\right]F(r).
\end{equation}

Therefore, substituting \eqref{dirac6} in \eqref{dirac5}, we obtain the following second-order differential equation for the spinor $F(r)$ (or better, for its two components)
\begin{equation}\label{dirac7}
\left[\frac{d^2}{dr^2}+\frac{1}{r}\frac{d}{dr}-\frac{\gamma^2_s}{\alpha^2 r^2}-(m_0\Omega)^2 r^2+E_s\right]F_s(r)=0,
\end{equation}
where we define
\begin{equation}\label{dirac8}
\gamma_s\equiv\left(m_j-\frac{s\alpha}{2}\right)\gtrless 0, \ \  E_s\equiv\frac{f^2(\varepsilon)E^2-m_0^2-g^2(\varepsilon)k^2_z}{g^2(\varepsilon)}+\frac{2m_0\sigma\Omega}{\alpha}\left(m_j+\frac{s\alpha}{2}\right), \ \ \Omega=\Omega_{eff}=\Omega(\varepsilon)\equiv\frac{\omega_c}{2g(\varepsilon)},
\end{equation}
being $\Omega\geq 0$ (since $\omega_c\geq 0$) an effective angular frequency or a ``rainbow frequency'', and the parameter $s=\pm 1$ (spinorial parameter) emerged from an eigenvalue equation given by $\sigma^3 F=\pm F=sF$ (i.e., $s$ also represents the eigenvalues of $\sigma^3$) \cite{Bakke2018,B1,Johnson}, where $s=+1$ is for the upper component and $s=-1$ is for the lower component. Furthermore, it is important to mention here that this parameter should not be confused with the spin quantum number, also symbolized by $s$ and whose value is $s=1/2$ (i.e., it describes a 1/2-spin particle) \cite{Griffiths}. However, as the spin magnetic quantum number is symbolized by $m_s=\pm 1/2=\uparrow \downarrow$ (spin up and down) \cite{Griffiths}, it implies that we can write this quantum number in terms of $s=\pm 1$ such as $m_s=s/2$; therefore, Eq. \eqref{dirac7} could also be written in terms of $m_s$ (making $s\to 2m_s$) \cite{Naseri}.

\section{Relativistic Landau levels\label{sec3}}

To solve exactly and analytically Eq. \eqref{dirac7}, it is necessary first to define a new independent variable (which is dimensionless). In particular, a suitable choice for this (due to the form of the equation) is given by $w=m_0\Omega r^2\geq 0$. In that way, Eq. \eqref{dirac7} becomes (a new second-order differential equation)
\begin{equation}\label{dirac9}
\left[w\frac{d^{2}}{dw^{2}}+\frac{d}{dw}-\frac{\gamma^{2}_s}{4\alpha^2w}-\frac{w}{4}+\frac{E_s}{4m_0\Omega}\right]F_s(w)=0.
\end{equation}

Now, it is necessary to analyze the asymptotic limit or behavior of Eq. \eqref{dirac9} for $w\to 0$ and $w\to\infty$ (this is possible due to the range of $r$). Once this is done, we have a regular solution to this equation given by
\begin{equation}\label{dirac10}
F_s(w)=C_sw^{\frac{\vert\gamma_s\vert}{2\alpha}}e^{-\frac{w}{2}}\bar{F}_s(w), \ \ (\vert\gamma_s\vert>0),
\end{equation}
where $C_s>0$ are normalization constants, $\bar{F}_s(\rho)$ are unknown functions to be determined, and $F_s(w)$ must satisfy the following boundary conditions to be a normalizable solution (finite or physically acceptable solution)
\begin{equation}\label{conditions} 
F_s(w\to 0)=F_s(w\to\infty)=0.
\end{equation}

So, substituting \eqref{dirac10} in \eqref{dirac9}, we have the following second-order differential equation for $\bar{F}_s(w)$
\begin{equation}\label{dirac11}
\left[w\frac{d^{2}}{dw^{2}}+(\vert\bar{\gamma}_s\vert-w)\frac{d}{dw}-\Bar{E}_s\right]\bar{F}_s(w)=0,
\end{equation}
where we define
\begin{equation}\label{define}
\vert\bar{\gamma}_s\vert\equiv\frac{\vert \gamma_s\vert}{\alpha}+1, \ \ \Bar{E}_s\equiv\frac{\vert\bar{\gamma}_s\vert}{2}-\frac{E_s}{4m_0\Omega}.
\end{equation}

According to the literature \cite{Oliveira3,GRG2024,Chen,Chen2017}, Eq. \eqref{dirac11} is the well-known generalized Laguerre equation, whose solutions are the generalized Laguerre polynomials, written as $\bar{F}_s(w)=L^{\frac{\vert \gamma_s\vert}{\alpha}}_n(w)$. So, for we have finite polynomials, it implies that the quantity $\Bar{E}_s$ must be equal to a negative integer: $\Bar{E}_s=-n$ (a quantization condition), where $n=n_r=0,1,2,\ldots$ is the (radial) quantum number. Therefore, from this condition with \eqref{dirac8}, we obtain as a result the following relativistic Landau levels (relativistic energy spectrum) for Dirac fermions in the cosmic string spacetime in the context of rainbow gravity
\begin{equation}\label{spectrum1}
E^\pm_{n,m_j,s}=\pm\frac{1}{f(\varepsilon)}\sqrt{m_0^2+g^2(\varepsilon)k^2_z+2g(\varepsilon)m_0\omega_c\left(n+\frac{1}{2}+\frac{\Big|m_j-\frac{s\alpha}{2}\Big|-\sigma\left(m_j+\frac{s\alpha}{2}\right)}{2\alpha}\right)},
\end{equation}
or in terms of $m_s$, in the form
\begin{equation}\label{spectrum2}
E^\pm_{n,m_j,m_s}=\pm\frac{1}{f(\varepsilon)}\sqrt{m_0^2+g^2(\varepsilon)k^2_z+g(\varepsilon)m_0\omega_c\left(2n+1-2\sigma m_s+\frac{\vert m_j-m_s\alpha\vert-\sigma(m_j-m_s\alpha)}{\alpha}\right)},
\end{equation}
where the positive sign ($+$) describes the positive-energy states or solutions, i.e., a particle with spin up ($m_s=1/2$) or down ($m_s=-1/2$) and a spectrum $E_{particle}=E^+>0$, and the negative sign ($-$) describes the negative-energy states or solutions, i.e., an antiparticle with spin up ($m_s=1/2$) or down ($m_s=-1/2$) and a spectrum $E_{antiparticle}=-E^-=\vert E^-\vert>0$ (a particle with negative energy is actually an antiparticle with positive energy) \cite{Greiner}, respectively. However, how both particle and antiparticle spectra are equal ($\vert E^+\vert=\vert E^-\vert$), implies that they are symmetric spectra (i.e., we have a symmetry about $E=\pm\sqrt{m_0^2+k^2_z}\neq 0$). In particular, this symmetry is lost (i.e., $\vert E^+\vert\neq \vert E^-\vert$) when an extra term (extra energy) appears outside the square root \cite{Oliveira3,GRG2024}. So, as we can see clearly in \eqref{spectrum2}, the spectrum is quantized in terms of quantum numbers $n$, $m_j$ and $m_s$ (and labeled through them), and explicitly depends on the rainbow functions $f(\varepsilon)$ and $g(\varepsilon)$ (with $\varepsilon=\varepsilon_{n,m_j,m_s}$), charge parameter $\sigma$ (describes a positively or negatively charged fermion), cyclotron frequency $\omega_c$ (is what gives rise to the Landau levels), topological parameter $\alpha$ (is what gives rise to the cosmic string), and on the square rest energy $m_0^2$ and square $z$-momentum $k_z^2$, respectively. Besides, the spectrum also does not have a well-defined degeneracy, that is, the function of the parameter $\alpha$ is to break the degeneracy (degree) of the spectrum, in other words, the presence of a cosmic string modifies the energy levels with respect to the quantum number $m_j$ and breaks the degeneracy of such levels \cite{Medeiros}. In fact, in the absence of the cosmic string ($\alpha=1$), the degeneracy of the spectrum can be finite or infinite depending on the values of $m_j$ and $m_s$ \cite{Oliveira3}. It is also important to mention that the spectrum \eqref{spectrum1} is very similar to the bosonic case for spinless or spin-0 particles, where the Landau levels were obtained using the Klein-Gordon equation \cite{Bezerra2019}. However, the main difference between them is with respect to the quantum number associated with the angular part of the solution of the equation (our case is a spinor while for the bosonic case is a scalar function).

On the other hand, it is also interesting to analyze the spectrum according to the values (sign) of the quantum number $m_j$ and charge parameter $\sigma$. Once this is done, we obtain Table \eqref{tab1}, which shows four possible settings for the spectrum. So, according to this table, we see that for $m_j>0$ (settings 1 and 2) the spectrum can or not depend on $m_j$ or $m_s$ (and $\alpha$), for example, for a positively charged fermion ($\sigma=+1$) the spectrum depends on $m_s$ but not on $m_j$ (and $\alpha$), while for a negatively charged fermion ($\sigma=-1$) the opposite happens, that is, the spectrum depends on $m_j$ (and $\alpha$) but not on $m_s$, respectively. In other words, the spectrum for a fermion with charge and positive angular momentum depends on both quantum numbers $n$ and $m_s$ (in this case, the spin affects the spectrum) while for a fermion with negative charge and positive angular momentum depends on both quantum numbers $n$ and $m_j$ (in this case, the spin does not affect the spectrum, that is, the spectrum is the same regardless of the chosen spin). Likewise, for a negative angular momentum $m_j<0$ (settings 3 and 4) something similar happens, for example, for a positively charged fermion ($\sigma=+1$) the spectrum depends on $m_j$ (and $\alpha$) but not on $m_s$, while for a negatively charged fermion ($\sigma=-1$) the opposite happens, that is, the spectrum depends on $m_s$ but not on $m_j$ (and $\alpha$), respectively. In other words, the spectrum for a fermion with charge and negative angular momentum depends on both quantum numbers $n$ and $m_s$ (in this case, the spin affects the spectrum) while for a fermion with positive charge and negative angular momentum depends on both quantum numbers $n$ and $m_j$ (in this case, the spin does not affect the spectrum). In addition, we also see that the spectrum of settings 2 and 3 are exactly the same, that is, the spectrum of a fermion with a negative charge and a positive angular momentum is the same as that of a fermion with a positive charge and a negative angular momentum. However, the spectrum of settings 1 and 4 are not the same (although both depend only on $n$ and $m_s$), that is, this happens because in one spectrum it appears $-2m_s$ while in the other it appears $+2m_s$. In this case, is as if the fermion ``lives in the spacetime without a cosmic string'' (the spectrum does not depend on $\alpha$).
\begin{table}[h]
\centering
\begin{small}
\caption{Relativistic spectrum depend on the values of $m_j$ and $\sigma$.} \label{tab1}
\begin{tabular}{ccc}
\hline
Setting & $(m_j,\sigma)$ & Spectrum \\
\hline
1& $(m_j>0,+1)$ & \ \ \ $E^\pm_{n,m_s}=\pm\frac{1}{f(\varepsilon)}\sqrt{m_0^2+g^2(\varepsilon)k^2_z+g(\varepsilon)m_0\omega_c\left(2n+1-2m_s\right)}$\\
2& $(m_j>0,-1)$ & \ \  \ \ $E^\pm_{n,m_j}=\pm\frac{1}{f(\varepsilon)}\sqrt{m_0^2+g^2(\varepsilon)k^2_z+g(\varepsilon)m_0\omega_c\left(2n+1+2\frac{m_j}{\alpha}\right)}$\\
3& $(m_j<0,+1)$ & \ \ \ \ \ \ $E^\pm_{n,m_j}=\pm\frac{1}{f(\varepsilon)}\sqrt{m_0^2+g^2(\varepsilon)k^2_z+g(\varepsilon)m_0\omega_c\left(2n+1+2\frac{\vert m_j\vert}{\alpha}\right)}$\\
4& $(m_j<0,-1)$ & \ \  \ $E^\pm_{n,m_s}=\pm\frac{1}{f(\varepsilon)}\sqrt{m_0^2+g^2(\varepsilon)k^2_z+g(\varepsilon)m_0\omega_c\left(2n+1+2m_s\right)}$\\
\hline
\end{tabular}
\end{small}
\end{table}

Before we graphically analyze the behavior of the energy levels (energy spectrum) for the three different rainbow gravity scenarios as a function of the magnetic field $B$ and of the parameter $\alpha$ for different values of $n$ (with $m_j=1/2$), let us first compare the spectrum \eqref{spectrum2}, or the spectra from Table \eqref{tab1} (for $m_j<0$), with some other works in the literature. So, we verified that in the absence of the cosmic string and of rainbow gravity ($\alpha=f(\varepsilon)=g(\varepsilon)=1$), we obtain the usual relativistic Landau levels in the Minkowski spacetime for $k_z\neq 0$ ($(3+1)D$) \cite{Gao,Johnson,Bermudez,Chen,Chen2017,Mameda,Miransky,Laurila,Naseri} and for $k_z=0$ ($(2+1)D$) \cite{Haldane,Schakel,Lamata,Miransky,Bermudez,Ishikawa,Jentschura,Bruckmann}. Therefore, we clearly see that our relativistic spectrum generalizes several particular cases in the literature.

Now, we will graphically analyze the behavior of energy levels, where we will use some information adopted in the bosonic case \cite{Bezerra2019}. In addition, as we also want the contribution of $\alpha$ in the spectrum, let us consider here, for example (and for the sake of convenience), the spectrum of setting $2$, which is the spectrum of the electron and positron. Then, using the three pairs of rainbow functions, given by \eqref{f1}, \eqref{f2}, and \eqref{f3}, on such a spectrum, we obtain Table \eqref{tab2}, which shows the spectrum for each one of the three scenarios (where $E_{n,m_j}=\vert E_{electron}\vert=\vert E_{positron}\vert>0$).
\begin{table}[h]
\centering
\begin{small}
\caption{Relativistic spectrum for each one of the three scenarios} \label{tab2}
\begin{tabular}{cc}
\hline
Scenario & Spectrum \\
\hline
First  & \ \ \ $E_{n,m_j}=\sqrt{m_0^2+(1-\eta\varepsilon)k^2_z+\sqrt{1-\eta\varepsilon}m_0\omega_c\left(2n+1+2\frac{m_j}{\alpha}\right)}$\\
Second & \ \ \ $E_{n,m_j}=\frac{\eta\varepsilon}{(e^{\eta\varepsilon}-1)}\sqrt{m_0^2+k^2_z+m_0\omega_c\left(2n+1+2\frac{m_j}{\alpha}\right)}$\\
Third  & \ \ \ $E_{n,m_j}=(1-\eta\varepsilon)\sqrt{m_0^2+\frac{k^2_z}{(1-\eta\varepsilon)^2}+\frac{m_0\omega_c}{(1-\eta\varepsilon)}\left(2n+1+2\frac{m_j}{\alpha}\right)}$\\
\hline
\end{tabular}
\end{small}
\end{table}

However, it is difficult to proceed without an adequate approximation (or simplification) of the spectra of Table \eqref{tab2}. According to Ref. \cite{Bezerra2019}, this approximation can be done through a Taylor series expansion up to the first order of the parameter $\eta$. Therefore, doing this, with $\varepsilon\to\varepsilon^0=E^0/E_P$, where $E^0=E^{f=g=1}$ is the spectrum with $f = g = 1$ (not to be confused with the limit \eqref{limit}), and writing the spectra as a function of the rainbow parameter $\varepsilon_{n,m_j}=E_{n,m_j}/E_P\geq 0$, we then obtain Table \eqref{tab3} with the approximate spectrum for each one of the three scenarios (and are even similar to the bosonic case \cite{Bezerra2019}).
\begin{table}[h]
\centering
\begin{small}
\caption{Approximate relativistic spectrum for each one of the three scenarios.} \label{tab3}
\begin{tabular}{cc}
\hline
Scenario & Spectrum \\
\hline
First  & \ \ \ $\varepsilon_{n,m_j}\approx\sqrt{\frac{m_0^2}{E^2_P}+\frac{k^2_z}{E^2_P}+\frac{m_0\omega_c}{E^2_P}\left(2n+1+2\frac{m_j}{\alpha}\right)}-\frac{\eta m_0\omega_c}{4E^2_P}\left(2n+1+2\frac{m_j}{\alpha}\right)-\frac{\eta k^2_z}{2E^2_P}$\\
Second & \ \ \ $\varepsilon_{n,m_j}\approx\sqrt{\frac{m_0^2}{E^2_P}+\frac{k^2_z}{E^2_P}+\frac{m_0\omega_c}{E^2_P}\left(2n+1+2\frac{m_j}{\alpha}\right)}-\frac{\eta m_0\omega_c}{2E^2_P}\left(2n+1+2\frac{m_j}{\alpha}\right)-\frac{\eta}{2}\left[\frac{m_0^2+k^2_z}{E^2_P}\right]$\\
Third  & \ \ \ $\varepsilon_{n,m_j}\approx\sqrt{\frac{m_0^2}{E^2_P}+\frac{k^2_z}{E^2_P}+\frac{m_0\omega_c}{E^2_P}\left(2n+1+2\frac{m_j}{\alpha}\right)}-\frac{\eta m_0\omega_c}{2E^2_P}\left(2n+1+2\frac{m_j}{\alpha}\right)-\frac{\eta m_0^2}{E^2_P}$\\
\hline
\end{tabular}
\end{small}
\end{table}

\begin{figure}[!h]
\centering
\includegraphics[width=18.1cm]{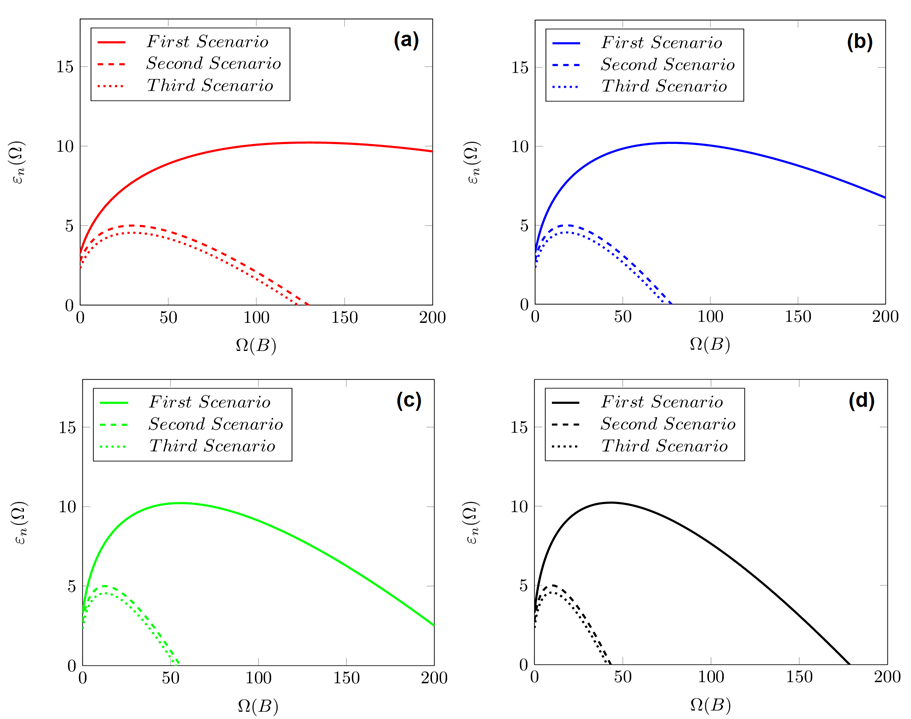}
\caption{Behavior of $\varepsilon_n(\Omega)$ versus $\Omega(B)$ for four different values of $n$.}
\label{fig1}
\end{figure}

According to Table \eqref{tab3}, we see that for $m^2_0=k^2_z$ the last two scenarios are exactly the same, so to prevent this, here we will consider that $m^2_0>k_z^2$ (i.e., the square rest energy of the fermion is greater than its square $z$-momentum). In addition, analogous to the bosonic case (or bosonic Landau levels) \cite{Bezerra2019}, here the Landau levels are also smaller relative to the Landau levels in the absence of rainbow gravity, i.e., here rainbow gravity also decreases the Landau level values.  In this way, we are now ready to graphically analyze the behavior of (dimensionless) energy levels. So, in Fig. \ref{fig1} we have the behavior (or graph) of $\varepsilon_n(\Omega)$ versus $\Omega(B)$ for four different values of $n$ with $m_j=1/2$, where (a) is for $n=0$ (ground state), (b) is for $n=1$ (first excited state), (c) is for $n=2$ (second excited state), (d) is for $n=3$ (third excited state), and we define the dimensional parameter $\Omega(B)$ in the form $\Omega(B)\equiv m_0\omega_c/E^2_P=e B/E^2_P\geq 0$ and we also consider for simplicity (in arbitrary units) that $m^2_0/E^2_P=10$, $k^2_z/E^2_P=1$, $\eta=0.1$ and $\alpha=1/2$.

According to Fig. \ref{fig1}, we see that the energies can increase or decrease as a function of $\Omega(B)$, that is, as the magnetic field increases, the energies increase to a maximum value and then decay tending to zero. For example, for the first scenario, the maximum energy is approximately $10$, while for the second and third scenarios (exact values), are $5$ and $4.55$, respectively. So, though the square rest energy is $10$ times the value of the square $z$-momentum ($m^2_0=10\times k^2_z$), the energies of the last two scenarios are very close to each other (and for $m^2_0=k^2_z$, the curves of both scenarios practically coincide). Furthermore, as the quantum number $n$ increases, the values of $\Omega$ (or the field) become smaller so that the energies tend to zero, i.e., the energies tend to zero more quickly as $n$ increases. In particular, these zero energies do not mean that the particle (or antiparticle) has no energy, but that the square root energy is equal to the energy outside this root (please see Table \eqref{tab3}). So, comparing the three scenarios, we clearly see that the energies are greater for the first scenario.

\begin{figure}[!h]
\centering
\includegraphics[width=18.1cm]{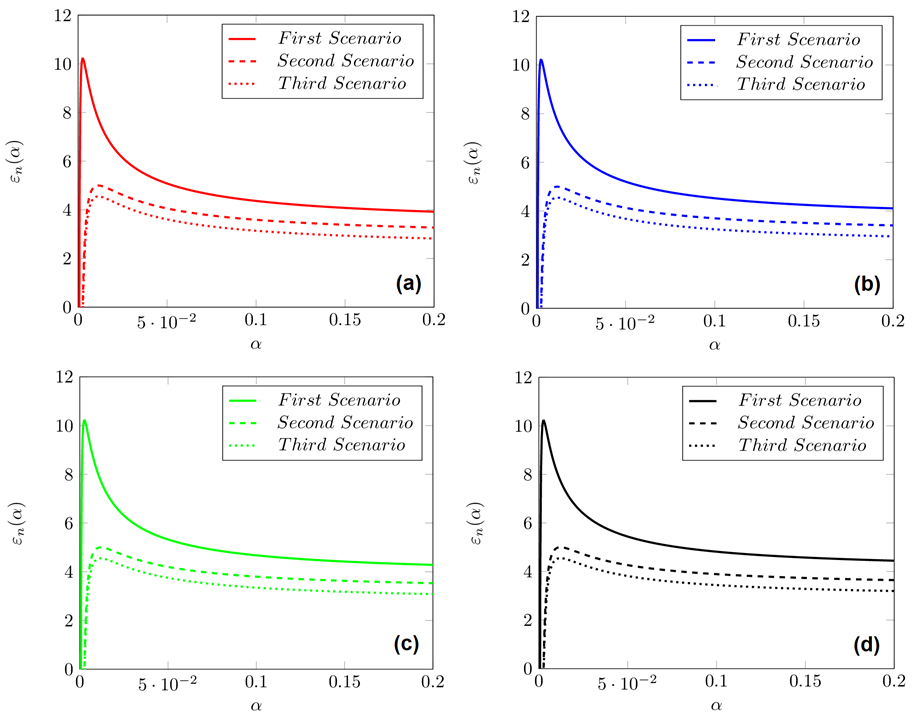}
\caption{Behavior of $\varepsilon_n(\alpha)$ versus $\alpha$ for four different values of $n$.}
\label{fig2}
\end{figure}

Now, let us graphically analyze the behavior of energy levels as a function of $\alpha$ (or of the conical curvature of the cosmic string). Therefore, in Fig. \ref{fig2} we have the behavior (or graph) of $\varepsilon_n (\alpha)$ versus $\alpha$ for four different values of $n$ with $m_j=1/2$, where (a) is for $n=0$, (b) is for $n=1$, (c) is for $n=2$, and (d) is for $n=3$, and we use $m^2_0/E^2_P=10$, $k^2_z/E^2_P=\Omega(B)=1$, and $\eta=0.1$. So, according to this figure, we see that the energies can increase or decrease as a function of $\alpha$, that is, as the curvature increases (which is $\alpha$ decreasing), the energies increase to a maximum value (approximately $10$ for the first scenario, exactly $5$ for the second scenario and approximately $4.5$ for the third scenario) and then decreases abruptly, tending to zero (i.e., the square root energy is equal to the energy outside this root). In a way, this is somewhat similar behavior to the previous figure (in the sense that it reaches a maximum and then tends to zero). However, although increasing $n$ does not change the maximum values of the energies (i.e., the maximum energies for each scenario are not changed by $n$), such a quantum number slightly increases the energies before they reach their respective maximum value. Furthermore, comparing the three scenarios, we clearly see that the energies are also greater for the first scenario (similar to the previous figures).

\section{Nonrelativistic Landau levels}\label{sec4}

Now, let us study the nonrelativistic limit (or low-energy limit) of our results. To achieve this, we can use a standard prescription often used in literature to get the nonrelativistic limit of a relativistic wave equation for massive particles (i.e., $m_0\neq 0$) \cite{Greiner,Lawrie}. In particular, this prescription considers that most of the total energy of the system is concentrated in the rest energy of the particle, that is, that $E_{R}=m_0+E_{NR}$, where it must satisfy: $m_0 \gg E_{NR}$. Furthermore, according to Ref. \cite{Bezerra2019}, for low energies the nonrelativistic wave equation (or the energy spectrum) must not contain the function $f(\varepsilon)$, that is, only the spatial part of the line element \eqref{lineelement2} should be considered, where now the rainbow parameter must be written as $\varepsilon_{NR}=E_{NR}/E_P$ (indeed, for a line element with $dt^2=0$ and positive signature, automatically results in the line element in Euclidean space). Therefore, applying all this information in \eqref{spectrum1}, or in \eqref{spectrum2}, with $E=E_R$, we obtain the following nonrelativistic Landau levels (nonrelativistic energy spectrum) for spin-1/2 particles (Pauli particles) in the cosmic string spacetime in the context of rainbow gravity
\begin{equation}\label{spectrum3}
E_{NR}=\frac{g^2(\varepsilon_{NR})k^2_z}{2m_0}+g(\varepsilon_{NR})\omega_c\left(n+\frac{1}{2}+\frac{\Big|m_j-\frac{s\alpha}{2}\Big|-\sigma\left(m_j+\frac{s\alpha}{2}\right)}{2\alpha}\right),
\end{equation}
or in terms of $m_s$, in the form
\begin{equation}\label{spectrum4}
E_{NR}=\frac{g^2(\varepsilon_{NR})k^2_z}{2m_0}+\frac{1}{2}g(\varepsilon_{NR})\omega_c\left(2n+1-2\sigma m_s+\frac{\vert m_j-m_s\alpha\vert-\sigma(m_j-m_s\alpha)}{\alpha}\right).
\end{equation}
 
So, through the spectrum above we obtain Table \eqref{tab4}, which shows four possible settings for the spectrum depending on the values of $m_j$ and $\sigma$.
\begin{table}[h]
\centering
\begin{small}
\caption{Nonrelativistic spectrum depend on the values of $m_j$ and $\sigma$.} \label{tab4}
\begin{tabular}{ccc}
\hline
Setting & $(m_j,\sigma)$ & Spectrum \\
\hline
1& $(m_j>0,+1)$ &  \ \  \ \ $E_{NR}=\frac{g^2(\varepsilon_{NR})k^2_z}{2m_0}+\frac{1}{2}g(\varepsilon)\omega_c\left(2n+1-2m_s\right)$\\
2& $(m_j>0,-1)$ &  \ \  \ \ $E_{NR}=\frac{g^2(\varepsilon_{NR})k^2_z}{2m_0}+\frac{1}{2}g(\varepsilon)\omega_c\left(2n+1+2\frac{m_j}{\alpha}\right)$\\
3& $(m_j<0,+1)$ &  \ \  \ \ \ \ \ $E_{NR}=\frac{g^2(\varepsilon_{NR})k^2_z}{2m_0}+\frac{1}{2}g(\varepsilon)\omega_c\left(2n+1+2\frac{\vert m_j\vert}{\alpha}\right)$\\
4& $(m_j<0,-1)$ &  \ \  \ \ $E_{NR}=\frac{g^2(\varepsilon_{NR})k^2_z}{2m_0}+\frac{1}{2}g(\varepsilon)\omega_c\left(2n+1+2m_s\right)$\\
\hline
\end{tabular}
\end{small}
\end{table}

In particular, we note that the spectrum \eqref{spectrum4}, or the spectra from Table \eqref{tab4}, has some similarities and differences with the relativistic case (or relativistic particle). For example, similar to the relativistic case, the nonrelativistic spectrum
\begin{itemize}
\item only admits positive-energy states or solutions ($E_{NR}\geq 0$), whose spectrum is also for a particle with spin-1/2, i.e., spin up ($m_s=+1/2$) and spin down ($m_s=-1/2$);
\item has its degeneration broken due to the parameter $\alpha$;
\item depends on $n$, $m_j$ and $\alpha$ only for $m_j>0$ with $\sigma=-1$, or for $m_j<0$ with $\sigma=+1$. In these cases, the spectrum does not depend on $m_s$;
\item depends on $n$ and $m_s$ only for $m_j<0$ with $\sigma=-1$, or for $m_j>0$ with $\sigma=+1$. In these cases, the spectrum does not depend on $m_j$ and $\alpha$ (It is as if the cosmic string did not exist);
\item requires a Taylor series expansion to be graphically analyzed (such as occurs in Ref. \cite{Bezerra2019} with the spectrum generated by the Schrödinger equation).
\end{itemize}

However, unlike the relativistic case, the nonrelativistic spectrum
\begin{itemize}
\item depends linearly on $k^2_z$, $\omega_c$, and $g(\varepsilon_{NR})$, as well as on $n$, $m_j$ and $m_s$;
\item does not depend on $f(\varepsilon_{NR})$.
\end{itemize}

Before we graphically analyze the behavior of the energy levels for the three different rainbow gravity scenarios as a function of the magnetic field $B$ and of the parameter $\alpha$ for different values of $n$ (with $m_j= 1/2$), let us first compare the spectrum \eqref{spectrum3} or \eqref{spectrum4}, or the spectra from Table \eqref{tab4}, with some other works in the literature (for particles with or without spin). For example, considering particles with spin (modeled by the Pauli Hamiltonian or Zeeman term), we verified that in the absence of the cosmic string and of rainbow gravity ($\alpha=g(\varepsilon_{NR})=1$), with $m_j<0$ and $\sigma=-1$ (setting 4), we obtain the Landau levels for $k_z=0$ \cite{Jentschura,Bermudez} and for $k_z\neq 0$ \cite{Gao}, and with $m_j>0$ and $\sigma=+1$ (setting 1), we obtain the Landau levels for $k_z\neq 0$ \cite{Vagner}. Now, considering spinless particles (modeled by the Schrödinger Hamiltonian), we verified that in the absence of the cosmic string and of rainbow gravity ($\alpha=g(\varepsilon_{NR})=1$), with $\sigma=-1$ and $\vert m_j-s/2\vert+(m_j+s/2)\to \vert l\vert+l$ ($l=0,\pm 1,\pm 2,\ldots$), we obtain the Landau levels for $k_z=0$ \cite{Ribeiro,Vagner,Wakamatsu1,Wakamatsu2,Li1999,Bhuiyan,Landau,Rosas,Schakel,Brand,Landry} and for $k_z\neq 0$ \cite{Mameda,Vagner,Wakamatsu2,Landau,Johnson}. Therefore, we clearly see that our nonrelativistic spectrum also generalizes several particular cases in the literature.

Then, using the three pairs of rainbow functions for $g(\varepsilon_{NR})$, given by \eqref{f1}, \eqref{f2}, and \eqref{f3}, on the spectrum of setting 2 of Table \eqref{tab4} together with a Taylor series expansion up to the first order of $\eta$, we obtain Table \ref{tab5}, which shows the approximate spectrum for each one of the three scenarios (actually for the first and second scenario) and written in terms of $\varepsilon_{NR}$. Here, it is important to mention that, unlike the relativistic case and the first scenario of Table \ref{tab5}, but analogous to the Landau levels for a nonrelativistic spinless particle \cite{Bezerra2019}, the Landau levels for the third scenario are higher relative to the Landau levels in the absence of rainbow gravity, i.e., rainbow gravity for the third scenario increases the Landau level values.
\begin{table}[h]
\centering
\begin{small}
\caption{Approximate nonrelativistic spectrum for each one of the three scenarios.} \label{tab5}
\begin{tabular}{cc}
\hline
Scenario & Spectrum \\
\hline
First  & \ \ \ $\varepsilon_{NR}\approx\left(\frac{k_z^2}{2m_0 E_P}+\frac{\omega_c}{2E_P}\left(2n+1+2\frac{m_j}{\alpha}\right)\right)\left(1-\frac{\eta}{2}\left(\frac{k_z^2}{m_0 E_P}+\frac{\omega_c}{2E_P}\left(2n+1+2\frac{m_j}{\alpha}\right)\right)\right)$\\
Second & \ \ \ $\varepsilon_{NR}=\frac{k_z^2}{2m_0 E_P}+\frac{\omega_c}{2E_P}\left(2n+1+2\frac{m_j}{\alpha}\right)$\\
Third  & \ \ \ $\varepsilon_{NR}\approx\left(\frac{k_z^2}{2m_0 E_P}+\frac{\omega_c}{2E_P}\left(2n+1+2\frac{m_j}{\alpha}\right)\right)\left(1+\eta\left(\frac{k_z^2}{m_0 E_P}+\frac{\omega_c}{2E_P}\left(2n+1+2\frac{m_j}{\alpha}\right)\right)\right)$\\
\hline
\end{tabular}
\end{small}
\end{table}

\begin{figure}[!h]
\centering
\includegraphics[width=18.1cm]{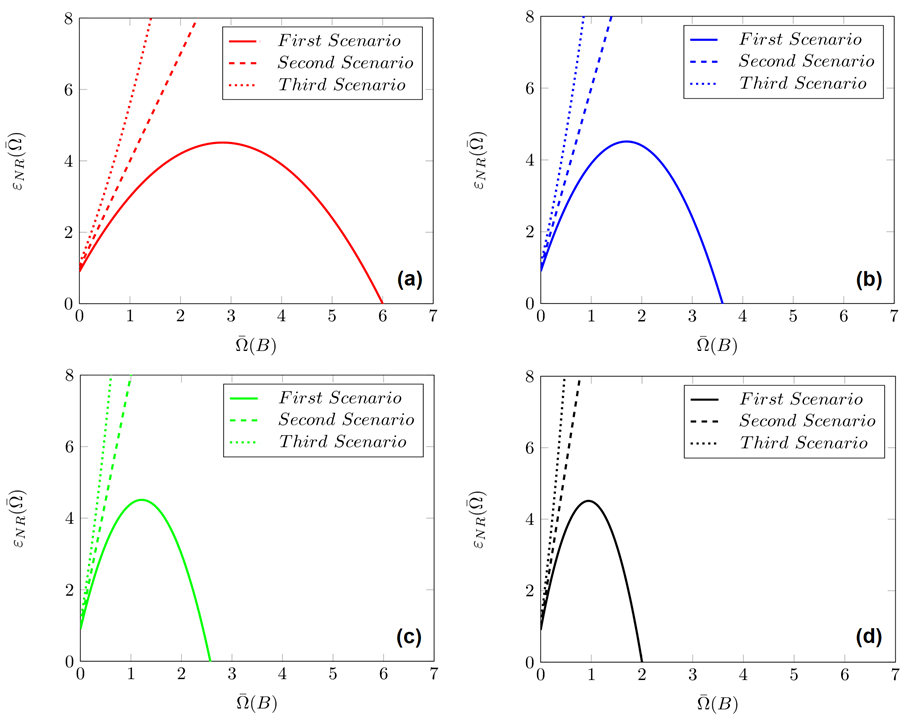}
\caption{Behavior of $\varepsilon_{NR}(\bar{\Omega})$ versus $\bar{\Omega}(B)$ for four different values of $n$.}
\label{fig3}
\end{figure}

So, in Fig. \ref{fig3} we have the behavior (or graph) of $\varepsilon_{NR}(\bar{\Omega})$ versus $\bar{\Omega}(B)$ for four different values of $n$ with $m_j=1/2$, where (a) is for $n=0$ (ground state), (b) is for $n=1$ (first excited state), (c) is for $n=2$ (second excited state), (d) is for $n=3$ (third excited state), and we define the dimensional parameter $\bar{\Omega}(B)$ in the form $\bar{\Omega}(B)\equiv\omega_c/2E_P=m_0 e B/2E_P\geq 0$ and we also consider for simplicity (in arbitrary units) that $k^2_z/2m_0E_P=1$, $\eta=0.1$ and $\alpha=1/2$. So, according to this figure, we see that the energies for the second and third scenarios always increase as a
function of $\Bar{\Omega}$ (i.e., the higher $\Bar{\Omega}$ the greater the energies), while for the first scenario can increase (whose maximum value is approximately $4.5$) or decrease (tending to zero). Besides, as the quantum number $n$ increases, the values of $\Bar{\Omega}$ become smaller so that the energies tend to zero, i.e., the energies tend to zero more quickly as $n$ increases. Now, comparing the three scenarios, we clearly see that the energies are greater for the third scenario.

\begin{figure}[!h]
\centering
\includegraphics[width=18.1cm]{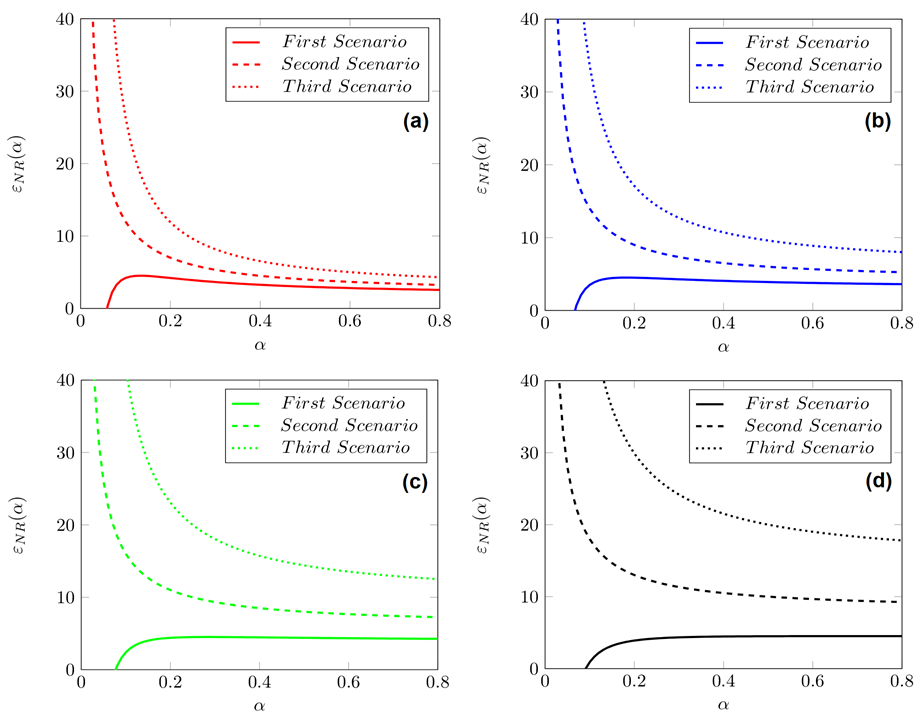}
\caption{Behavior of $\varepsilon_{NR}(\alpha)$ versus $\alpha$ for four different values of $n$.}
\label{fig4}
\end{figure}

Already in Fig. \ref{fig4}, we have the behavior (or graph) of $\varepsilon_{NR}(\alpha)$ versus $\alpha$ for four different values of $n$ with $m_j=1/2$, where (a) is for $n=0$, (b) is for $n=1$, (c) is for $n=2$, and (d) is for $n=3$, and we use $k^2_z/2m_0E_P=\bar{\Omega}(B)=1$, and $\eta=0.1$. So, according to this figure (which has somewhat similar behavior to the previous figure), we see that the energies for the second and third scenarios always increase as a function of $\alpha$ (i.e., the smaller $\alpha$ the greater the energy), while for the first scenario can be (approximately) constants (for $0.2<\alpha\leq 1$) or decrease (tending to zero for $\alpha<0.2$). Besides, the energies of the three scenarios increase with increasing $n$. Now, comparing the three scenarios, we clearly see that the energies are also greater for the third scenario.

\section{Conclusions}\label{sec5}

In this paper, we study the relativistic and nonrelativistic Landau levels for Dirac fermions in the cosmic string spacetime in the context of rainbow gravity in $(3+1)$-dimensions, i.e., we study the fermionic case or version of Ref. \cite{Bezerra2019} (in which Klein-Gordon scalar bosons were studied). To carry out this study, we work with the curved Dirac equation with minimal coupling in cylindrical coordinates, where the formalism used was the tetrads formalism of GR. In fact, we chose this formalism because it is considered an excellent tool for studying fermions in curved spacetimes. With respect to the line element (or metric) of our problem, we use the line element of the (static) cosmic string spacetime modified by rainbow gravity, where the cosmic string is described by a parameter $\alpha$ (depends on the mass of the cosmic string) and rainbow gravity is described by two real functions $f(\varepsilon)$ and $g(\varepsilon)$, being $\varepsilon=E/E_P$ a positive real parameter (rainbow parameter), where $E$ is the relativistic total energy of the fermion and $E_P$ is the Planck energy. In particular, we consider three different pairs of rainbow functions (or three different scenarios) for $f(\varepsilon)$ and $g(\varepsilon)$, whose choice is based on phenomenological motivations and also because are the most worked in the literature. So, these rainbow functions are: $f(\varepsilon)=1$ and $g(\varepsilon)=\sqrt{1-\eta\varepsilon}$ (first scenario); $f(\varepsilon)=\frac{e^{\eta\varepsilon}-1}{\eta\varepsilon}$ and $g(\varepsilon)=1$ (second scenario); and $f(\varepsilon)=g(\varepsilon)=\frac{1}{1-\eta\varepsilon}$ (third scenario), where $\eta$ is a positive real parameter.

So, to solve exactly and analytically a second-order differential equation generated by the Dirac equation, we had to follow two steps, that is, define a new independent variable, given by $w=m_0\Omega r^2$, and then analyze the asymptotic behavior of the new second-order differential equation for $w\to 0$ and $w\to\infty$. Done this, we obtain a generalized Laguerre equation, where the last term of this equation obeys a quantization condition (i.e., the last term must be equal to a negative integer). Consequently, we obtain from this condition the relativistic Landau levels (relativistic energy spectrum) for a fermion/particle (positive-energy states) and an antifermion/antiparticle (negative-energy states) in the cosmic string spacetime in the context of rainbow gravity. In particular, we verified that this spectrum is symmetric (particle and antiparticle have the same energy), quantized in terms of quantum numbers $n$, $m_j$ and $m_s$, where $n=0,1,2,\ldots$ is the radial quantum number, $m_j=\pm 1/2,\pm 3/2,\ldots$ is the total magnetic quantum number, $m_s=s/2=\pm 1/2$ is the spin magnetic quantum number, and explicitly depends on the rainbow functions $f(\varepsilon)$ and $g(\varepsilon)$, charge parameter $\sigma=\pm 1$ (describes a positively or negatively charged fermion), cyclotron frequency $\omega_c=eB/m_0$ (is what gives rise to the Landau levels), being $B$ the strength of the external uniform magnetic field, curvature or topological parameter $\alpha$, and on the square rest energy $m^2_0$ and square $z$-momentum $k^2_z$, respectively. Besides, the spectrum does not have a well-defined degeneracy, i.e., the function of $\alpha$ is to break the degeneracy of the spectrum.

On the other hand, analyzing the spectrum according to the values of $m_j$ and $\sigma$, we obtain four possible settings for the spectrum. For example, we see that for $m_j>0$ and $\sigma=+1$, or $m_j<0$ and $\sigma=-1$, the spectrum depends on $n$ and $m_s$ but not on $m_j$ and $\alpha$ (in this case, is as if the fermion ``lives in the spacetime without a cosmic string''), while for $m_j>0$ and $\sigma=-1$, or $m_j<0$ and $\sigma=+1$, the spectrum depends on $n$ and $m_j$ but not on $m_s$ (in this case, the spin does not affect the spectrum, that is, the spectrum is the same regardless of the spin). Before we graphically analyze the behavior of the energy levels (or spectrum) for the three scenarios as a function of $B$ and $\alpha$ for four values of $n$ (with $m_j=1/2$), first we compare the spectrum with some other works in the literature. For example, we verified that in the absence of the cosmic string and of rainbow gravity ($\alpha=f(\varepsilon)=g(\varepsilon)=1$), we
obtain the usual relativistic Landau levels in the Minkowski spacetime for $k_z\neq 0$ and $k_z=0$, i.e., our relativistic spectrum generalizes several particular cases in the literature. So, to analyze the behavior of the spectrum, we use a Taylor series expansion up to the first order of $\eta$, where we obtain an approximate spectrum for each scenario. In that way, in the graph ``Energy versus Magnetic Field'', we see that the energies can increase or decrease as a function of $B$, that is, the energies increase to a maximum value and then decay tending to zero. Furthermore, as the quantum number $n$ increases, the values of $B$ become smaller so that the energies tend to zero (i.e., the energies tend to zero more quickly as $n$ increases). However, comparing the three scenarios, we see that the energies are greater for the first scenario. Already in the graph ``Energy versus curvature'', we see that the energies can increase or decrease as a function of $\alpha$, that is, the energies increase to a maximum value and then decrease abruptly, tending to zero. Besides, although increasing $n$ does not change the maximum energies, such a quantum number slightly increases the energies before they reach their respective maximum values. However, comparing the three scenarios, we see that the energies are also greater for the first scenario.

Lastly, we also study the nonrelativistic limit of our results. For example, considering that most of the total energy of the system stays concentrated in the rest energy of the particle, we obtain the nonrelativistic Landau levels (nonrelativistic energy spectrum) for spin-1/2 particles (Pauli particles) in the cosmic string spacetime in the context of rainbow gravity. In particular, we note that this spectrum has some similarities and differences with the relativistic case, i.e., similar to the relativistic case, the nonrelativistic spectrum only admits positive-energy states; has its degeneration broken due to the parameter $\alpha$; depends on $n$, $m_j$ and $\alpha$ only for $m_j>0$ with $\sigma=-1$, or $m_j<0$ with $\sigma=+1$; depends on $n$ and $m_s$ and only for $m_j>0$ with $\sigma=+1$, or $m_j<0$ with $\sigma=-1$; and requires a Taylor series expansion to be graphically analyzed. Now, unlike the relativistic case, the nonrelativistic spectrum depends linearly on $k^2_z$, $\omega_c$, and $g(\varepsilon_{NR})$ as well as on $n$, $m_j$ and $m_s$; and does not depend on $f(\varepsilon_{NR})$. Besides, we verified that in the absence of the cosmic string and of rainbow gravity ($\alpha=g(\varepsilon_{NR})=1$), we
obtain the usual nonrelativistic Landau levels for $k_z\neq 0$ and $k_z=0$, i.e., our nonrelativistic spectrum also generalizes several particular cases in the literature. So, regarding the graphs for this low energy regime, we see that in the graph ``Energy versus Magnetic Field'' the energies for the second and third scenarios always increase as a function of $B$ while for the first scenario can increase (up to a maximum value) or decrease (tending to zero). However, comparing the three scenarios, we see that the energies are greater for the third scenario. Already in the graph ``Energy versus curvature'', we see that the energies for the second and third scenarios always increase as a function of $\alpha$ while for the first scenario can be (approximately) constants or decrease (tending to zero). However, comparing the three scenarios, we see that the energies are also greater for the third scenario.

\section*{Acknowledgments}

\hspace{0.5cm}The author would like to thank the Conselho Nacional de Desenvolvimento Cient\'{\i}fico e Tecnol\'{o}gico (CNPq) for financial support.

\section*{Data availability statement}

\hspace{0.5cm} This manuscript has no associated data or the data will not be deposited. [Author’s comment: There is no data because this is theoretical work based on calculations to obtain the relativistic and nonrelativistic Landau levels for Dirac fermions in the cosmic string spacetime in the context of rainbow gravity.]

\end{document}